\documentclass[twocolumn,showpacs,amsmath,amssymb,prl]{revtex4}
\usepackage{graphicx,amstext}
\usepackage{dcolumn}
\usepackage{bm}
\newcommand{\vs}{{\it vs.\@}}

\newcommand{\al}{{\it et al.\@}}

\newcommand{\sr}{Sr$_{14-x}$Ca$_{x}$Cu$_{24}$O$_{41}$}
\newcommand{\bq}{\begin{equation}}
\newcommand{\eq}{\end{equation}}

\begin{document}

\title{Suppression of the charge-density-wave state in Sr$_{14}$Cu$_{24}$O$_{41}$ by calcium doping}

\author{T. Vuleti\'{c}}
 \email{tvuletic@ifs.hr}
\author{B. Korin-Hamzi\'{c}}
\author{S. Tomi\'{c}}
 \homepage{http://www.ifs.hr/real_science}
\affiliation{Institut za fiziku, P.O. Box 304, HR-10001 Zagreb, Croatia}
\author{B. Gorshunov}
 \altaffiliation {Permanent address: General Physics Institute, Russian Academy of Sciences, Moscow, Russia.}
\author{P. Haas}
\author{T. ~R\~o\~om}
 \altaffiliation {Permanent address: National Institute of Chemical Physics and Biophysics, Tallinn, Estonia.}
\author{M. Dressel}
\affiliation{1.\@ Physikalisches Institut, Universit\"{a}t Stuttgart, D-70550
Stuttgart, Germany}
\author{J. Akimitsu}
\author{T. Sasaki}
\author{T. Nagata}
\affiliation{Department of Physics, Aoyama-Gakuin University, Tokyo, Japan}

\date{\today}

\begin{abstract}
The charge response in the spin chain/ladder compound \sr~is characterized by DC resistivity,
low-frequency dielectric spectroscopy and optical spectroscopy. We identify a phase transition below which
a charge-density wave (CDW) develops in the ladder arrays. Calcium doping suppresses this phase with the
transition temperature decreasing from 210 K for $x=0$ to 10 K for $x=9$, and the CDW gap from 130 meV
down to 3 meV, respectively. This suppression is due to the worsened nesting originating from the increase of the inter-ladder tight-binding hopping integrals, as well as from disorder introduced at the Sr sites. These results altogether speak in favor of two-dimensional superconductivity under pressure.
\end{abstract}

\maketitle

The discovery of superconductivity under pressure in the two-leg ladder compound \sr~has provoked much
attention since it is the first superconducting copper oxide material with a non-square-lattice layered
pattern~\cite{Uehara96,Nagata97}. Theoretically, in an undoped spin ladder system, spin singlets in rungs
produce the ground state, usually referred to as a gapped spin-liquid~\cite{Takigawa98}. Upon Ca doping,
the added charge carriers (holes) tend to share the same rung on the ladder in order to minimize the
energy paid to break the spin singlets. The formation of hole pairs is expected to lead to
superconductivity with d-wave symmetry, whereas the spin gap remains the same or decreases equally in
size. Due to the quasi one-dimensional nature of ladders, a competing charge-ordered state (charge-density
wave), with hole pairs as building entities, may prevent the occurrence of superconductivity. The ladder
systems provide, therefore, a nice possibility to study spin and charge dynamics, and their interplay in a
spin-gapped environment with signi\-fi\-cance to the phase diagram of high-temperature superconducting
cuprates.

\sr~is such a quasi-one-dimensional system but shows additional complexity since it contains a plane of
Cu$_2$O$_3$ ladders, a plane of one-dimensional CuO$_2$ chains and a (Sr,Ca) layer. These three distinct
layers are placed in the crystallographic a-c plane and stack alternatingly along the b-axis. The spin
gaps are found to induce the activated temperature dependence of the spin-lattice NMR relaxation rate and
the Knight shift in both the CuO$_2$ chains and the Cu$_2$O$_3$ ladders below about 50 K and 200 K,
respectively~\cite{Takahashi97,Kumagai97}. While the spin gap for the chain $\Delta_{\rm spin}^{\rm
chains}\approx14$~meV remains constant on doping, the spin gap for the ladder $\Delta_{\rm spin}^{\rm
ladders}\approx50$~meV decreases linearly with the Ca content. As far as the charge order is concerned,
its existence is well established in the CuO$_2$ chain sub-unit below 200 K as a result of the
antiferromagnetic dimers pattern. The charge gap induced by this order leads to a sharp peak observed in
magnetic Raman-scattering measurements~\cite{Gozar01,Schmidt}.

No definite understanding has been reached yet on the nature of the charge-ordered state in the ladders
(the subsystem responsible for the conductivity and superconductivity of the compound) and its evolution
on calcium-doping, which leads to superconductivity. Along the most conducting direction (c-axis), a
semiconductor-like temperature dependence of the DC resistivity is observed,  which does not change
appreciably upon doping, while a metallic behavior is found only for $x\geq11$ and $T >
50$~K~\cite{Motoyama97}. The low-frequency dielectric and the optical response of the parent compound
Sr$_{14}$Cu$_{24}$O$_{41}$ reveals a broad relaxation centered in the radio-frequency range with an
Arrhenius-like decay determined by the resistive dissipation~\cite{Gorshunov02}. We have shown that this
relaxation is due to screened phason excitations of the charge-density wave (CDW), which develops in the
ladders and produces a CDW pinned mode at microwave frequencies, thus determining the low-temperature
insulating ground state in pure Sr$_{14}$Cu$_{24}$O$_{41}$. Our findings were confirmed by Blumberg
\al~\cite{Blumberg02}.

In order to address the relationship between CDW, gapped spin-liquid, and superconducting states, and thus
to get an insight on the mechanism of superconductivity in spin ladders, we have performed systematic
investigation of how the CDW instability in  Sr$_{14}$Cu$_{24}$O$_{41}$ evolves upon Ca doping, which is
essential to achieve superconductivity. We find that (i) the phase transition to the CDW state is strongly
suppressed by calcium doping, revealing a competition between the charge-density wave and
superconductivity; and (ii) a complex relation exists between CDW and gapped spin-liquid ground states.

The DC resistivity was measured between 2 K and 700 K. In the frequency range 0.01 Hz - 1 MHz the spectra
of the dielectric function were obtained from the complex conductance measured at 2 K$ < T <$200
K~\cite{Pinteric01}. These spectra were complemented by reflectivity measurements at 8 - 10000 cm$^{-1}$
using quasi-optical~\cite{Kozlov98} and infrared spectrometers, from which the spectra of the complex
dielectric function were obtained by a Kramers-Kronig analysis. All measurements were done along the
crystallographic c-axis of high-quality single  crystals.
\begin{figure}
\centering\includegraphics[clip,scale=0.36]{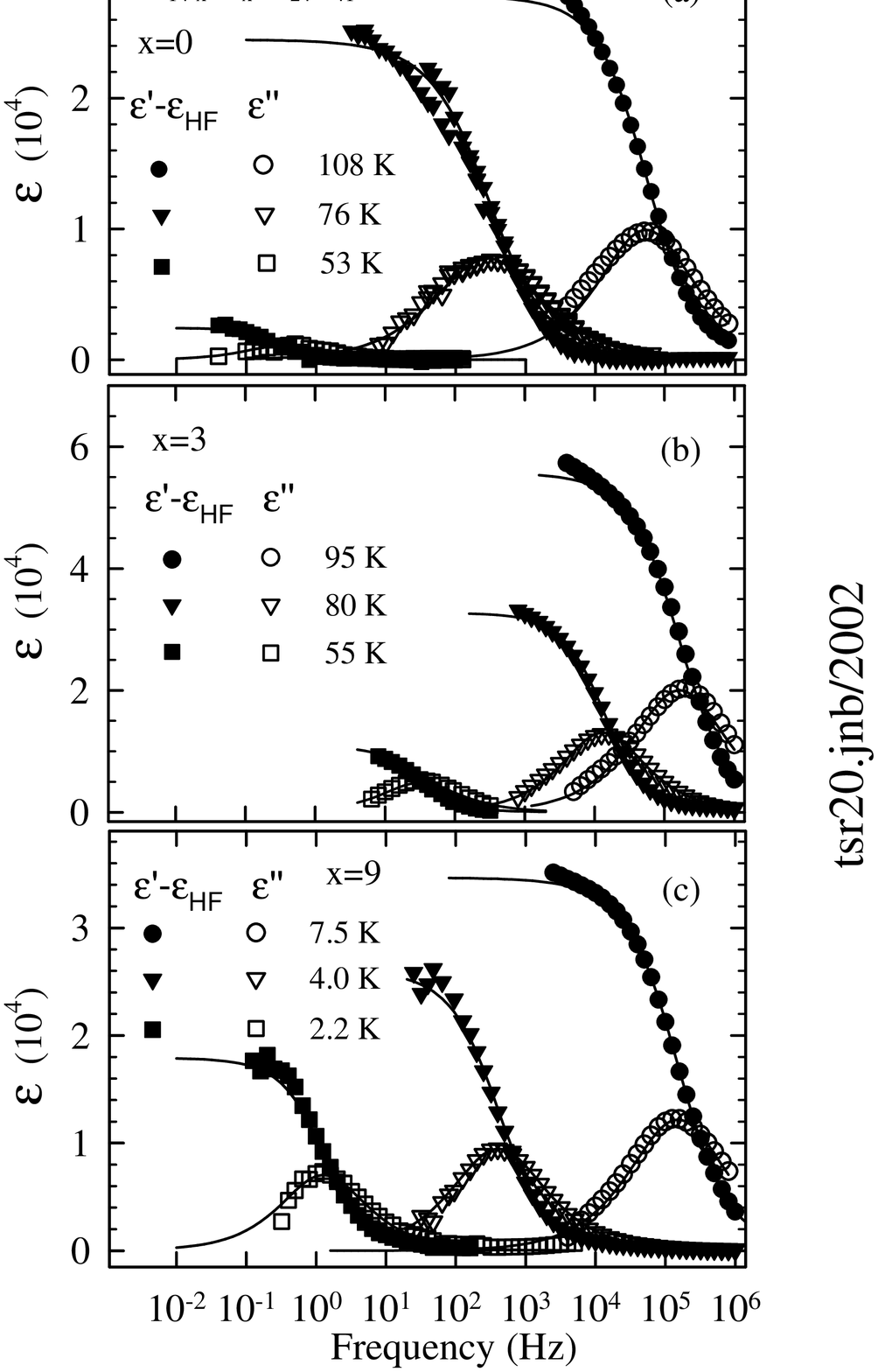} 
\caption{
Real and imaginary parts of the dielectric
function of \sr~for (a) $x = 0$, (b)  $x = 3$ and (c) $x = 9$ measured at three representative
temperatures as a function of frequency with the AC electric field applied along the c-axis. The full
lines are from fits by the generalized Debye expression: $\varepsilon(\omega) -
\varepsilon_{HF}=\Delta\varepsilon/[1+(\imath\omega\tau_0)^{1-\alpha}]$.} \label{Fig1}
\end{figure}

\begin{figure}
\centering\includegraphics[clip,scale=0.41]{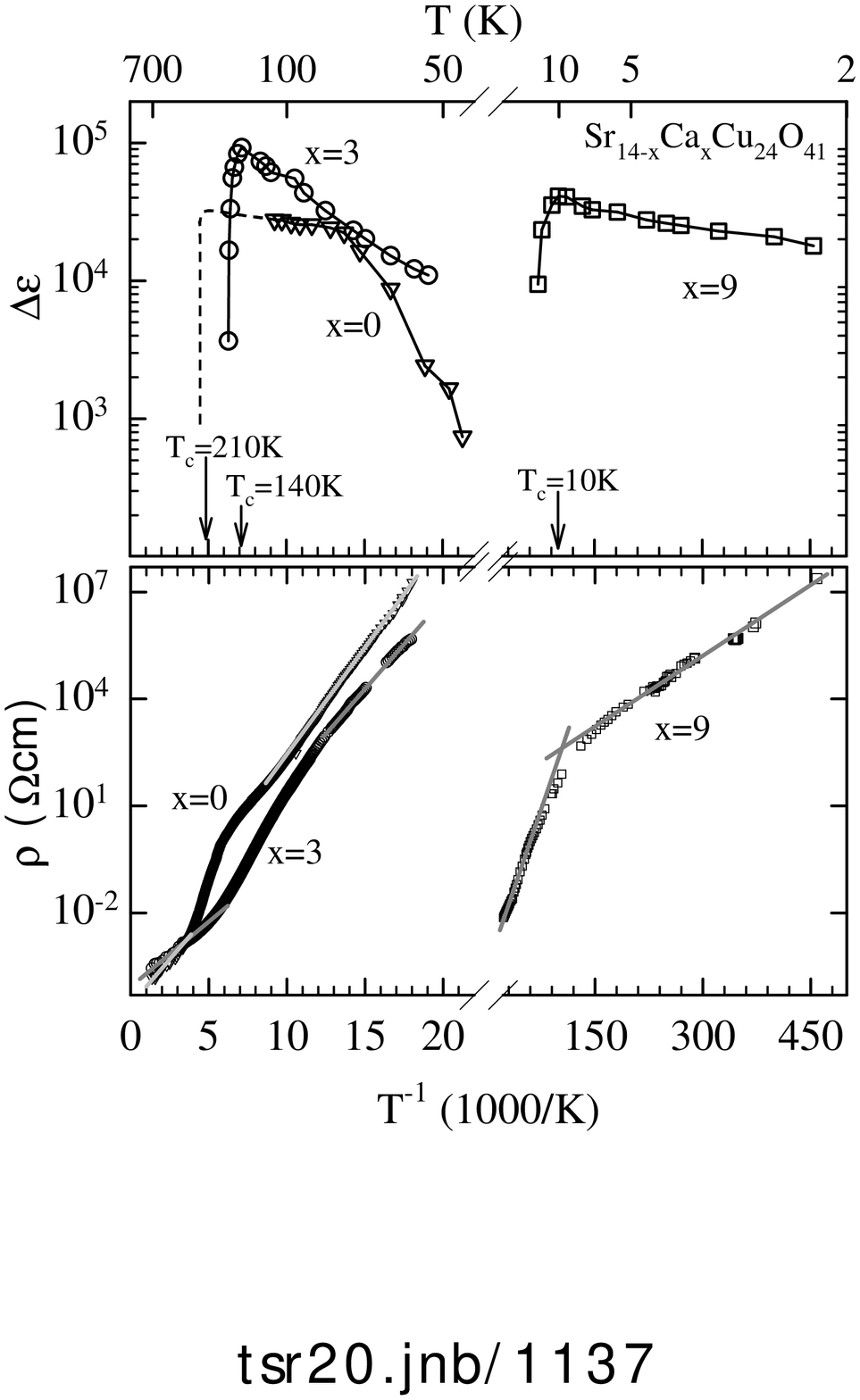} 
\caption{ 
Temperature dependence of the dielectric strength $\Delta\varepsilon$ of the radio-frequency CDW-related mode (upper panel) and DC resistivity $\rho$ (lower panel) in \sr~for three different Ca concentrations $x$. The arrows indicate the CDW phase transition temperature $T_c$. In the upper panel the full lines guide the eye, while in the lower panel the lines represent fits by Arrhenius functions. An assumed behavior of $\Delta\varepsilon$ for $x = 0$, based on that observed for $x = 3$ and 9, is represented by a dashed line.
} \label{Fig2}
\end{figure}
Fig. \ref{Fig1} shows the frequency dependent complex dielectric response at three selected temperatures
for $x = 0, 3$ and 9. A pronounced dielectric relaxation is observed for all three compositions providing
evidence for the CDW formation~\cite{Gorshunov02}. The screened loss peak ($\varepsilon$'') centered at
$\tau_0^{-1}$ moves toward lower frequencies and smaller amplitudes with decreasing temperature. The main
features of this relaxation do not qualitatively change on doping: the dielectric strength
$\Delta\varepsilon=\varepsilon_0 - \varepsilon_{HF} \approx 5\times10^4$ ($\varepsilon_0$ and
$\varepsilon_{HF}$ are static and high-frequency dielectric constants), the symmetric broadening of the
relaxation-time distribution given by $1-\alpha \approx 0.8$, and the mean relaxation time $\tau_0$, which
closely follows a thermally activated behavior in a manner similar to the DC resistivity [$\tau_0(T)
\propto \rho(T)$]. Our results clearly demonstrate a distinct phase transition to the CDW ground state,
which is strongly suppressed by Ca-doping (Fig. \ref{Fig2}): on decreasing temperature, a sharp growth of
$\Delta\varepsilon$ starts in the close vicinity of $T_c$ and reaches the huge value of the order of $10^4
- 10^5$ at $T_c = 140$~K and $T_c = 10$~K for $x = 3$ and 9, respectively. These $T_c$ values perfectly
correspond to the temperature of the phase transition as determined in the DC resistivity measurements,
indicated by pronounced peaks at $T_c$ in the derivative of the resistivity~\cite{Gorshunov03}. The
overall decrease of $\Delta\varepsilon$  below $T_c$  for $x = 0$ is substantial, while it becomes much
less pronounced for $x = 3$ and $x = 9$. While for $x = 3$ and 9 we were able to track the dielectric mode
into the respective phase transition region, for $x = 0$ the mode leaves the upper bound (1 MHz) of our
frequency window already at $T > 110$~K.

\begin{figure}
\centering\includegraphics[clip,scale=0.40]{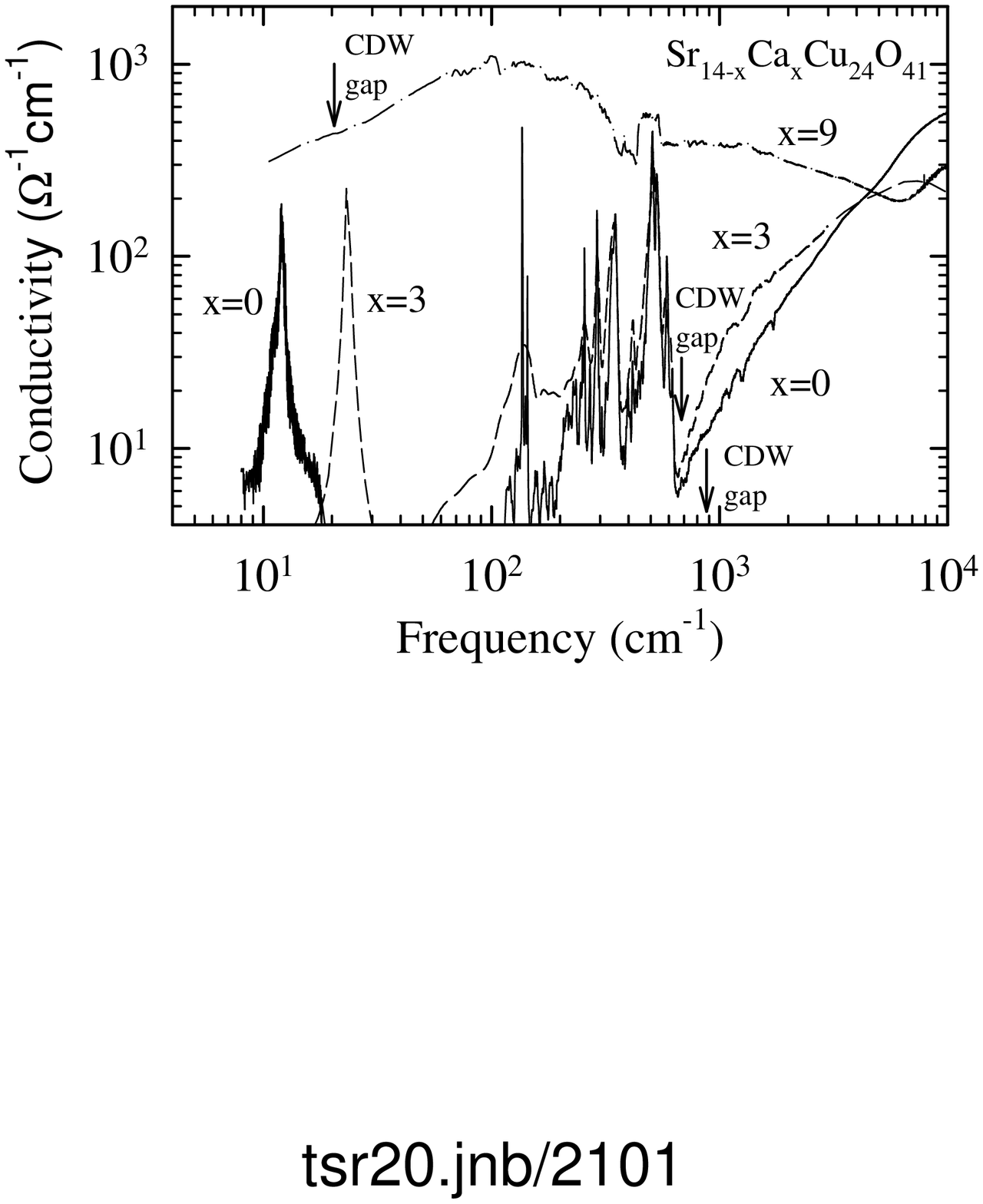} 
\caption{ 
Low-temperature ($T=5$~K) optical conductivity of \sr~for $x=0, 3, 9$. The spectra are obtained for the electric field vector of radiation $E||c$. The arrows show the CDW gaps obtained from activated DC resistivity.} 
\label{Fig3}
\end{figure}

Fig.\ref{Fig3} shows the low-temperature spectra of the optical conductivity for three different Ca
compositions, $x = 0, 3$ and 9. The optical conductivity decreases below approximately 2500 cm$^{-1}$ for
$x = 0$ and 3, and around 25 cm$^{-1}$ for $x = 9$, which we associate with the opening of the CDW gap.
This decrease is only observed for $x = 0$, $x = 3$ and $x = 9$ at temperatures lower than
240~K, 150~K and 20~K, respectively~\cite{Gorshunov02,Gorshunov03}. These temperatures nicely agree with the CDW phase
transition temperature $T_c$ as determined from our DC and low-frequency dielectric measurements (Fig.
\ref{Fig2}),~\cite{Gorshunov03}. Moreover, the CDW gap values extracted from these measurements (130 meV
$\approx$ 900 cm$^{-1}$, 110 meV $\approx$ 750 cm$^{-1}$ and 3 meV $\approx$ 20 cm$^{-1}$, for $x =0$, 3
and 9, respectively) correspond well to the edges seen in the optical conductivity spectra. We thus
conclude that the broad peak-like feature around 100 cm$^{-1}$ for $x = 9$, which appears only for
$T\leq20$~K, cannot be assigned to the pinned CDW mode as suggested by Osafune \al~\cite{Osafune99},
rather it can simply be attributed to the opening of the CDW gap. High AC conductivity, as compared with the DC value, which remains at frequencies lower than the CDW gap, is due to a power law dispersion.

\begin{figure}
\centering\includegraphics[clip,scale=0.41]{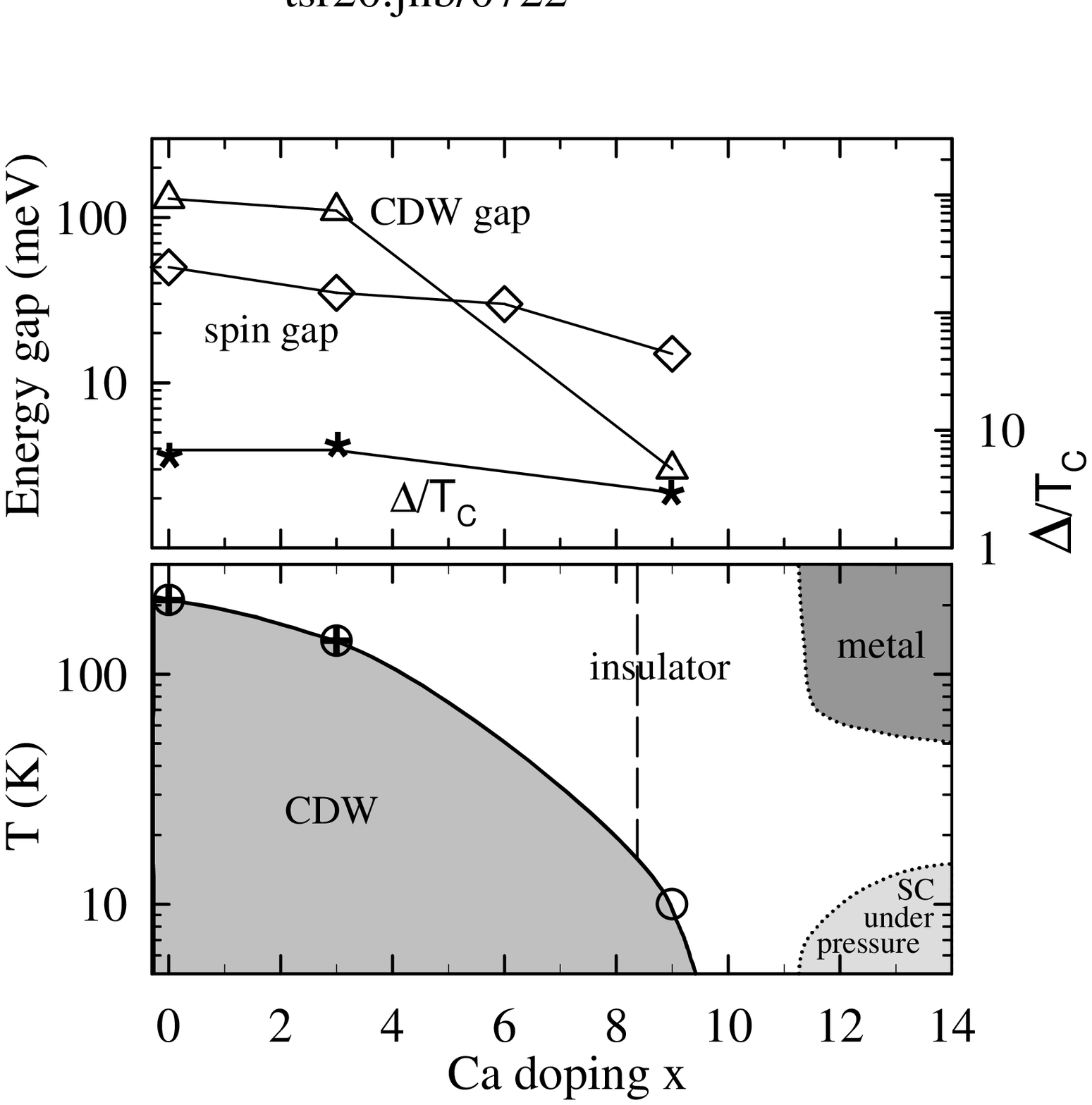} 
\caption{ 
Upper panel: Dependences of the CDW related parameters in \sr~on the calcium concentration: CDW gap $\Delta$  (triangles) and the ``strength'' of the CDW interaction $\Delta/T_c$ (stars). The dependence of the spin gap (diamonds) in the ladders is also shown~\cite{Kumagai97}. Lower panel: Qualitative phase diagram for \sr~as a function of Ca-doping. The CDW phase is determined by the CDW critical temperature $T_c$ (open circles), which coincides with the paramagnetic -- spin-gapped phase crossover-temperature $T^*$ (crosses)~\cite{Imai98}. Also sketched is the insulating, the metallic~\cite{Motoyama97}, and the superconducting phase, which appears under pressure. The activation energy in the insulating phase equals the CDW gap on the left of the dashed line, while it is close to the spin gap of the ladders on the right side.}
\label{Fig4}
\end{figure}
Fig.\ref{Fig4} demonstrates the doping dependence of the CDW-related parameters: the CDW gap $\Delta$,
the critical temperature $T_c$ and $\Delta/T_c$, which is a measure of the interaction strength leading to
the CDW formation. A strong decrease of the CDW phase transition temperature and the CDW single-particle
gap induced by Ca-doping is striking. The broadening of the phase transition (as seen from the logarithmic
derivative of resistivity \vs~inverse temperature~\cite{Gorshunov03}) and the decrease of $T_c$, might be
attributed to a disorder introduced by Ca-doping at Sr sites. In order to explain such a pronounced
effect, we suggest that in addition to the above, the CDW ground state is suppressed because the warping
of the quasi one-dimensional band, which is associated with the ladder sub-units, increases with
Ca-doping. Indeed, a more two-dimensional hole distribution in the ladders has been revealed recently by
X-ray absorption spectroscopy as the major effect of the Ca substitution~\cite{Nucker00}. The "strength"
of the phase transition $\Delta/T_c$ is much larger than the mean-field $\Delta/T_c = 3.5$ for $x = 0$ and
3, and drops to about 3.5 for $x = 9$. In the quasi-1D CDW systems, $\Delta/T_c$ is usually found to be
enhanced above the mean-field value and ascribed to strong 1D correlations. The observed drop for $x = 9$
may be due to the increased dimensionality induced at high Ca-doping levels.

Now we address the question of the nature and the origin of the observed CDW ground state. The peak we
previously assigned to the pinned phason mode~\cite{Gorshunov02} centered at 12 cm$^{-1}$ for $x = 0$,
shifts to 23 cm$^{-1}$ for $x = 3$ and eventually disappears for $x = 9$ (see Fig. \ref{Fig3}). In our
recent optical measurements on La$_{3}$Sr$_{3}$Ca$_{8}$Cu$_{24}$O$_{41}$ we see the same peak centered
at about 12 cm$^{-1}$~\cite{Vuletic03} although there is no hole transfer into the conducting
ladders~\cite{Nucker00, Osafune97}. Hence we conclude that the 12 cm$^{-1}$ peak is not a signature of the
CDW phason mode~\cite{footnote2}.

Therefore, we follow the assignment of the peak at 1.8~cm$^{-1}$ to the CDW phason pinned mode, proposed
by Kitano \al~\cite{Kitano01}. The mode was detected for $x = 0$ and 3 doping, but not for $x =
9$~\cite{Kitano00}, which may be due to the screening by the large number of free carriers at this high
doping level. Since we observe the screened temperature-dependent response in the radio-frequency range
centered at $\tau_0^{-1}$ (see Fig. \ref{Fig1}), which represents a fingerprint of the CDW phason
response~\cite{Littlewood87}, for $x =0$ all the way up to $x = 9$, we can safely assume that the pinned
mode always exists around $\Omega_0 \approx 1.8$~cm$^{-1}$. Littlewood~\cite{Littlewood87} developed an
expression, which connects the microwave unscreened phason mode at $\Omega_0$ with the screened loss peak
in the radio-frequency range at $\tau_0^{-1}$:
$$
m^*=\frac{{\mathrm e}^2 n}{ \sigma_{DC}  \tau_0 m_0 \Omega_0^2}
$$
With $\sigma_{DC}=1/\rho$ and $n$ the carrier concentration condensed into the CDW, we can estimate the
effective mass $m^*$ of the condensate. We have suggested that the CDW forms in the ladder sub-unit because the
ladders rather than the chains represent the conducting channel in \sr~\cite{Gorshunov02}. In
La$_{3}$Sr$_{3}$Ca$_{8}$Cu$_{24}$O$_{41}$, which contains no holes in the ladders, we did not find any
signature of the CDW-related dielectric response; a strong confirmation of our
assumption~\cite{Vuletic03}. If only one out of six holes condenses into the CDW on the ladders, we
estimate the CDW effective mass to be in the range $20<m^*<50$; with basically no dependence on
temperature or Ca content for $x$ between 0 and 9. Note that the values for the CDW effective mass are
usually in the range $10^2< m^*<10^4$~\cite{Gruner88}. The small values clearly indicate that
electron-phonon interactions are weak. Therefore, a superstructure, which accompanies the CDW, is expected
to be weak and difficult to observe by X-ray diffuse scattering. Besides the small value of $m^*$, the CDW
state differs from those commonly observed in one-dimensional conductors~\cite{Gruner88} by the small
nonlinearity of resistivity with no finite threshold field~\cite{Gorshunov02,Kitano01,footnote3}, and a
strongly temperature dependent $\Delta\varepsilon$, which may be connected to a redistribution
of the holes between chains and ladders and an interplay between spins and charges~\cite{Gorshunov03}.

Finally, we comment on the complex relation of charge-density wave and gapped spin-liquid ground states.
The CDW phase transition temperature $T_c$, from the high-temperature (HT) phase into the CDW ground
state,  coincides with the crossover temperature $T^*$, which separates the high temperature paramagnetic
regime from the spin-gapped ground state in the ladders~\cite{Imai98}, (see Fig. \ref{Fig4}). A similar
gradual $x$-dependence is also seen by both the CDW gap and the spin gap. However, this seemingly direct
interdependence turns over completely at high Ca-doping levels, where the CDW gap drops to 3 meV, which is
5 times smaller than the spin gap~\cite{Kumagai97}.

This result supports the scenario according to which superconductivity is established in the presence of a
finite spin gap at doping levels where the competing CDW order fully vanishes. Pressure suppresses an
insulating HT phase, which still persists at these doping levels. Namely, the activation energy for low
Ca-contents ($x = 0, 3$) in the HT phase is very close to the CDW gap in the CDW ground state, while for
high $x = 9$ it drops substantially, but clearly less than the CDW gap, down to a size of 12 meV, close to
the value of spin gap in ladders. This means that the ambient pressure HT insulating  state at high doping
levels is different in nature as compared to the one at low doping, as indicated by the dashed line in
Fig. \ref{Fig4}. The origin of the HT insulating phase is not clear yet and might indicate strong electron
correlations effects. The role of applied pressure is then to diminish these correlations, to drive the
system away from the CDW instability by increasing further its dimensionality from 1 to 2, and setting
finally the SC state, which is thus of a two-dimensional nature.

We demonstrated the phase transition from a high-temperature insulating phase to the
charge-density wave ground state of \sr~and the gradual suppression of the CDW by
Ca-substitution of Sr. We propose that at high doping levels the CDW order fully vanishes due to
increased two-dimensionality and disorder, while a finite spin gap is still present. External pressure
then suppresses an insulating high temperature phase and establishes superconductivity.

We thank G. Untereiner for the preparation of the samples. This work was supported by the Croatian
Ministry of Science and Technology and the Deutsche Forschungsgemeinschaft (DFG).


\begin{thebibliography}{10}
\bibitem{Uehara96}
M. Uehara \al,
\newblock{ J.Phys.Soc.Japan} {\bf65}, 2764 (1996).
\bibitem{Nagata97}
T. Nagata \al,
\newblock{  Physica C} {\bf282-287}, 153 (1997).
\bibitem{Takigawa98}
M. Takigawa \al,
\newblock{ Phys.Rev.B} {\bf57}, 1124 (1998).
\bibitem{Takahashi97}
T. Takahashi \al,
\newblock{ Phys.Rev.B} {\bf56}, 7870 (1997).
\bibitem{Kumagai97}
K. Kumagai \al,
\newblock{ Phys.Rev.Lett.} {\bf78}, 1992 (1997).
\bibitem{Gozar01}
A. Gozar \al,
\newblock{ Phys.Rev.Lett.} {\bf87}, 197202 (2001).
\bibitem{Schmidt}
K.P. Schmidt \al,
\newblock{ Phys.Rev.Lett.} {\bf90}, 167201 (2003).
\bibitem{Motoyama97}
N. Motoyama \al,
\newblock{ Phys.Rev.B} {\bf55}, R3386 (1997).
\bibitem{Gorshunov02}
B. Gorshunov \al,
\newblock{ Phys.Rev.B} {\bf66}, 060508(R) (2002).
\bibitem{Blumberg02}
G. Blumberg \al,
\newblock { Science} {\bf297}, 584 (2002).
\bibitem{Pinteric01}
M. Pinteri\'{c} \al,
\newblock{ Eur. Phys. J. B} {\bf 22}, 335 (2001).
\bibitem{Kozlov98}
G. Kozlov and A. Volkov,  in: {\em Millimeter and Submillimeter Wave Spectroscopy of Solids},
ed.~G.Gr\"{u}ner (Springer-Verlag Berlin, 1998).
\bibitem{Gorshunov03}
B. Gorshunov \al,
\newblock{ to be published}
\bibitem{Osafune99}
T. Osafune \al,
\newblock{ Phys.Rev.Lett.} {\bf82}, 1313 (1999).
\bibitem{Nucker00}
N. N\"{u}cker \al,
\newblock{ Phys.Rev.B} {\bf62}, 14384 (2000).
\bibitem{Imai98}
T. Imai \al,
\newblock{ Phys.Rev.Lett.} {\bf81}, 220 (1998).
\bibitem{Vuletic03}
T.Vuleti\'{c} \al,
\newblock{ Phys.Rev.B} {\bf67}, 2145XX (2003) and cond-mat/0212439
\bibitem{Osafune97}
T. Osafune \al,
\newblock{ Phys.Rev.Lett.} {\bf78}, 1980 (1997).

\bibitem{footnote2}The
12 cm$^{-1}$ peak should rather be attributed to a phonon. This is supported by the fact that our
low-frequency dielectric spectroscopy measurements on La$_{3}$Sr$_{3}$Ca$_{8}$Cu$_{24}$O$_{41}$ show no
sign of the dielectric relaxation associated with CDW screened response.

\bibitem{Kitano01}
H. Kitano \al,
\newblock { Europhys. Lett.} {\bf56}, 434 (2001).
\bibitem{Kitano00}
H. Kitano \al,
\newblock{  Physica C} {\bf341-348}, 463 (2000).
\bibitem{Littlewood87}
P.B. Littlewood,
\newblock { Phys. Rev. B} {\bf36}, 3108 (1987).
\bibitem{Gruner88}
G. Gr\"{u}ner,
\newblock { Rev.Mod.Phys.} {\bf60}, 1129, 1988.
\bibitem{footnote3}
We have only, in contrast to  Blumberg \al~\cite{Blumberg02}, observed a negligibly small non-linear
conductivity for $x=0$, 3 and 9, which emerges from the noise background.
\end{thebibliography}
\end{document}